\documentclass[submission,copyright,creativecommons]{eptcs}
\providecommand{\event}{ACL2 2015} 
\usepackage{breakurl}              

\usepackage{amssymb}
\usepackage{amsmath}

\title{Perfect Numbers in ACL2}
\author{John Cowles \qquad\qquad Ruben Gamboa
 \institute{Department of Computer Science \\
            University of Wyoming \\
            Laramie, Wyoming, USA}
 \email{cowles@uwyo.edu \qquad\qquad ruben@uwyo.edu}
}

\def\titlerunning{Perfect Numbers in ACL2}
\def\authorrunning{J. Cowles \& R. Gamboa}

\newtheorem{thm}{Theorem}

\begin{document}

\maketitle

\begin{abstract}
A \textbf{perfect number} is a positive integer $n$ such that $n$ equals the sum of all positive integer
divisors of $n$ that are less than $n$. That is, although $n$ is a divisor of $n$, $n$ is excluded from
this sum. Thus $6 = 1 + 2 + 3$ is perfect, but $12 \neq 1 + 2 + 3 + 4 + 6$ is not perfect. An ACL2
theory of perfect numbers is developed and used to prove, in ACL2(r), this bit of mathematical folklore:
Even if there are infinitely many perfect numbers, the series below, of the reciprocals of all perfect numbers, converges.
            \[ \sum_{\mbox{\scriptsize{perfect}} \,  n} \frac{1}{n} \]               
\end{abstract}

\section{Perfect Numbers}

The smallest perfect numbers are $6 = 2 \cdot 3 = 2^{1} (2^{2} - 1)$,
$28 = 4 \cdot 7 = 2^{2} (2^{3} - 1)$, $496 = 16 \cdot 31 = 2^{4} (2^{5} - 1)$,
$8128 = 64 \cdot 127 = 2^{6} (2^{7} - 1)$. In each of these examples, the second factor, $3, 7, 31, 127$,
of the form $2^{k} - 1$, is a prime. The Greek Euclid proved~\cite[page 3]{dunham}:
\begin{thm}
If $2^{k} - 1$ is prime, then $n = 2^{k-1}(2^{k} - 1)$ is perfect.
\end{thm}

Primes of the form $2^{k} - 1$ are called \textbf{Mersenne primes}. Thus every new Mersenne prime
leads to a new perfect number. According to Wikipedia~\cite{wiki}, less than 50 Mersenne primes are known.
The largest known Mersenne prime is $2^{57,885,161} - 1$, making $2^{57,885,160}(2^{57,885,161} - 1)$
the largest known perfect number, with over $34$ million digits. It is not known if there are infinitely
many Mersenne primes, nor if there are infinitely many perfect numbers.

All perfect numbers built from Mersenne primes are even.
The Swiss Euler proved every \textbf{even} perfect number is built from some Mersenne prime~\cite[page 10]{dunham}: 
\begin{thm}
If $n$ is an even perfect number, then $n = 2^{k-1}(2^{k} - 1)$, where $2^{k} - 1$ is prime. 
\end{thm}

It is not known if there are any odd perfect numbers, but Euler also proved~\cite[page 250]{pollack}:
\begin{thm}
If $n$ is an odd perfect number, then $n = p^{i} m^{2}$, where $p$ is prime and $i$, $p$, $m$ are odd. 
\end{thm}
ACL2 is used to verify each of these three theorems.

If there are only finitely many perfect numbers, then clearly the series
             \[ \sum_{\mbox{\scriptsize{perfect}} \,  n} \frac{1}{n} \]
converges.  ACL2(r) is used to verify that even if there are \textbf{infinitely many perfect numbers},
the series converges.

\section{The ACL2 Theory}

In number theory, for positive integer $n$, $\sigma(n)$ denotes the sum of \textbf{all} (including $n$)
positive integer divisors of $n$. The function $\sigma(n)$ has many useful properties, so the definition
of a perfect number is reformulated in terms of $\sigma$~\cite[pages 8--9]{dunham}:
          \[ \mbox{perfect}(n) \ \ \mbox{if and only if} \ \ \sigma(n) = 2n \].

These six properties of $\sigma$ are among those formulated and proved in ACL2:
\begin{enumerate}
  \item $p$ is prime if and only if $\sigma(p) = p + 1$.
  \item If $p$ is prime, then $\sigma(p^{k}) = \sum_{i = 0}^{k} p^{i} = \frac{p^{k+1} - 1}{p - 1}$.
  \item If $p$ and $q$ are different primes, then $\sigma(p \cdot q) = \sigma(p) \cdot \sigma(q)$.
  \item $\sigma(k \cdot n) \le \sigma(k) \cdot \sigma(n)$
  \item If $\gcd(k,n) = 1$, then $\sigma(k \cdot n) = \sigma(k) \cdot \sigma(n)$.
  \item If $p$ is prime, then $\gcd(p^{k}, \sigma(p^{k})) = 1$.
\end{enumerate}

If $n = 2^{i}(2^{i+1}-1)$ is an even perfect number, then the exponent $i$ is computed by an ACL2 term,
\verb+(cdr (odd-2^i n))+, that returns the largest value of $i$ such that $2^{i}$ divides $n$.

If $n = p^{i} m^{2}$ is an odd perfect number, then $p, i, m$ are respectively computed by the ACL2 terms
\begin{itemize}
 \item \begin{verbatim}
(car (find-pair-with-odd-cdr
      (prime-power-factors n)))
       \end{verbatim}

  \item \begin{verbatim}
(cdr (find-pair-with-odd-cdr
      (prime-power-factors n)))
       \end{verbatim}
  \item \begin{verbatim}
(product-pair-lst (pairlis$ (strip-cars 
                             (remove1-equal 
                              (find-pair-with-odd-cdr
                               (prime-power-factors n))
                              (prime-power-factors n)))
                            (map-nbr-product 
                             1/2 
                             (strip-cdrs 
                              (remove1-equal 
                               (find-pair-with-odd-cdr
                                (prime-power-factors n))
                               (prime-power-factors n))))))
           \end{verbatim}
\end{itemize}
These terms implement the following computation:
\begin{enumerate}
 \item Factor $n = \prod_{j=0}^{k} p_{j}^{e_{J}}$ into the product of powers of distinct odd primes.
 \item Exactly one of the exponents, say  $e_{0}$, will be odd and all the other exponents will be even.
 \item $p$ is the prime with the odd exponent and $i$ is the unique odd exponent. So
       $n = p^{i} \cdot \prod_{j=1}^{k} p_{j}^{2f_{j}}$.
 \item Then $m = \prod_{j=1}^{k} p_{j}^{f_{j}}$ and $n = p^{i} m^{2}$.
\end{enumerate}

ACL2 is used to verify a result of B. Hornfeck, that different odd perfect numbers, 
$n_{1} = p_{1}^{i_{1}} m_{1}^{2} \ne n_{2} = p_{2}^{i_{2}} m_{2}^{2}$ have distinct $m_{i}$~\cite[page 251]{pollack}: 
\begin{thm}
If $n_{1} = p_{1}^{i_{1}} m_{1}^{2}$ and $n_{2} = p_{2}^{i_{2}} m_{2}^{2}$ are odd perfect numbers and $m_{1} = m_{2}$,
then $n_{1} = n_{2}$.
\end{thm}
Theorems 2, 3, and 4 are enough to prove the folklore that the series, of the reciprocals of all 
perfect numbers, converges.

\section{ACL2(r)}

ACL2(r)~\cite{gamboa} is based on \textbf{Nonstandard Analysis}~\cite{robinson,nelson} which provides 
rigorous foundations for reasoning about real, complex, infinitesimal, and infinite quantities. 
There are two versions of the \textbf{reals}
\begin{enumerate}
   \item The \textbf{Standard Reals}, $ ^{\mbox{st}}\mathbb{R} $, is the unique \textbf{complete}
         ordered field.  This means that
         \begin{itemize}
            \item  Every nonempty subset of  $ ^{\mbox{st}}\mathbb{R} $
                   that is bounded above has a \textbf{least upper bound}.
         \end{itemize}
         There are no non-zero infinitesimal elements, nor are there are any infinite elements in
         $ ^{\mbox{st}}\mathbb{R} $.
   \item The \textbf{HyperReals}, $ ^{\star}\mathbb{R} $, is a proper field extension of
         $ ^{\mbox{st}}\mathbb{R} $:
                $ ^{\mbox{st}}\mathbb{R} \varsubsetneqq  {^{\star}\mathbb{R}}$.  
         There are non-zero infinitesimal elements and also infinite elements in $ ^{\star}\mathbb{R} $. 
\end{enumerate}

Here are some technical definitions.
\begin{itemize}
 \item $x \in {^{\star}\mathbb{R}}$ is \textbf{infinitesimal}:
         For all positive $r \in {^{\mbox{st}}\mathbb{R}}$, $(|x| < r)$.

         $0$ is the only infinitesimal in ${^{\mbox{st}}\mathbb{R}}$.

         \texttt{(i-small x)} in ACL2(r).
 \item $x \in {^{\star}\mathbb{R}}$ is \textbf{finite}: 
         For some $r \in {^{\mbox{st}}\mathbb{R}}$, $(|x| < r)$.

          \texttt{(i-limited x)} in ACL2(r).
 \item $x \in {^{\star}\mathbb{R}}$ is \textbf{infinite}:  
         For all $r \in {^{\mbox{st}}\mathbb{R}}$, $(|x| > r)$.

         \texttt{(i-large x)} in ACL2(r) 
 \item $x, y \in {^{\star}\mathbb{R}}$ are \textbf{infinitely close}, $x \approx y$: 
         $x - y$ is infinitesimal.

         \texttt{(i-close x y)} in ACL2(r).
 \item $\mathfrak{n}_{\infty}$ is an infinite positive integer constant.

         \texttt{(i-large-integer)} in ACL2(r). 
\end{itemize}

Every (partial) function
$ f : {^{\mbox{st}}\mathbb{R}^n} \longmapsto {^{\mbox{st}}\mathbb{R}^k} $
has an extension  
$ {^{\star}\!f} : {^{\star}\mathbb{R}^n} \longmapsto {^{\star}\mathbb{R}^k} $
such that 
\begin{enumerate}
  \item For $x_{1}, \cdots, x_{n} \in {^{\mbox{st}}\mathbb{R}}$,
         ${^{\star}\!f}(x_{1}, \cdots, x_{n}) = f(x_{1}, \cdots, x_{n})$.
  \item Every first-order statement about $f$ true in ${^{\mbox{st}}\mathbb{R}}$
        is true about ${^{\star}\!f}$ in ${^{\star}\mathbb{R}}$.

        Example. \\
                  $(\forall x) [\sin^{2}(x) + \cos^{2}(x) = 1]$ is true in
                  ${^{\mbox{st}}\mathbb{R}}$.

                  $(\forall x) [{^{\star}\!\sin}^{2}(x) + {^{\star}\!\cos}^{2}(x) = 1]$ is true in
                  ${^{\star}\mathbb{R}}$. 
\end{enumerate}

Any (partial) function
$ f : {^{\mbox{st}}\mathbb{R}^n} \longmapsto {^{\mbox{st}}\mathbb{R}^k} $
is said to be \textbf{classical}. 
\begin{itemize}
     \item Identify a classical $f$ with its extension ${^{\star}\!f}$.

            That is, use $f$ for both the original classical function $f$ and
            its extension  ${^{\star}\!f}$.
     \item Use $(\forall^{\mbox{st}} x)$ for $(\forall x \in {^{\mbox{st}}\mathbb{R}})$,
            i.e. ``for all \textbf{standard} $x$.''

           Use $(\exists^{\mbox{st}} x)$ for $(\exists x \in {^{\mbox{st}}\mathbb{R}})$,
            i.e. ``there is some \textbf{standard} $x$.''
     \item ``$(\forall x) [\sin^{2}(x) + \cos^{2}(x) = 1]$ is true in
                     ${^{\mbox{st}}\mathbb{R}}$'' becomes 
                     ``$(\forall^{\mbox{st}} x)[\sin^{2}(x) + \cos^{2}(x) = 1]$
                     is true in ${^{\star}\mathbb{R}}$.''

           ``$(\forall x) [{^{\star}\!\sin}^{2}(x) + {^{\star}\!\cos}^{2}(x) = 1]$ is true in
                    ${^{\star}\mathbb{R}}$'' becomes
                    ``$(\forall x) [\sin^{2}(x) + \cos^{2}(x) = 1]$ is true in
                    ${^{\star}\mathbb{R}}$.'' 
    \end{itemize}
Numeric constants, $c$, are viewed as $0$-ary functions,
$ c : {^{\mbox{st}}\mathbb{R}^{0}} \longmapsto {^{\mbox{st}}\mathbb{R}} $
or $ c : {^{\star}\mathbb{R}^{0}} \longmapsto {^{\star}\mathbb{R}} $.
Thus, elements of ${^{\mbox{st}}\mathbb{R}}$, such as $2, 4, -1$, are classical.
But elements of ${^{\star}\mathbb{R}} - {^{\mbox{st}}\mathbb{R}}$, such as the infinite positive integer
$\mathfrak{n}_{\infty}$, are not classical.
Functions defined using the nonstandard concepts of
infinitesimal, finite, infinite, and infinitely close are not classical.

Let $f$ be a (partial) unary function, whose domain includes the \textbf{nonnegative} integers, into the reals.
Here are three possible definitions for the real series $\sum_{i=0}^{\infty} f(i)$ converges.
The first two are versions of Weierstrass' traditional definition that the real series converges.
One version for the standard reals, another version for the hyperreals.
\begin{enumerate}
   \item
\begin{verbatim}
(defun-sk
 Series-Converges-Traditional-Standard ( )
\end{verbatim}
     \vspace*{\smallskipamount}
  \qquad $(\exists^{\mbox{st}} L)
           (\forall^{\mbox{st}} \epsilon > 0)
             (\exists^{\mbox{st}} \; \mbox{integer} \; M > 0)
                   (\forall^{\mbox{st}} \; \mbox{integer} \; n)           
                      (n > M \Rightarrow |\sum_{i=0}^{n} f(i) - L| < \epsilon)$ \\
\verb+)+
   \item
\begin{verbatim}
(defun-sk
 Series-Converges-Traditional-Hyper ( )
\end{verbatim}
     \vspace*{\smallskipamount}
  \qquad $(\exists L)
           (\forall \epsilon > 0)
             (\exists \; \mbox{integer} \; M > 0)
                   (\forall \; \mbox{integer} \; n)           
                     (n > M \Rightarrow |\sum_{i=0}^{n} f(i) - L| < \epsilon)$ \\
\verb+)+
   \item
\begin{verbatim}
(defun-sk
 Series-Converges-Infinitesimal ( )
\end{verbatim}
     \vspace*{\smallskipamount}
  \qquad $(\exists^{\mbox{st}} L)
           (\forall \; \mbox{infinite integer}  \; n > 0)           
                      (\sum_{i=0}^{n} f(i) \approx L)$ \\
\verb+)+
\end{enumerate}
For \textbf{classical} $f$, ACL2(r) verifies these three definitions are equivalent.
ACL2(r) also verifies for classical $f$, with \textbf{nonnegative} range, these definitions are equivalent
to this nonstandard definition~\cite{cowles}:
\begin{itemize}
   \item
\begin{verbatim}
(defun
 Series-Converges-Nonstandard ( )
\end{verbatim}
     \vspace*{\smallskipamount}   
  \qquad $\sum_{i=0}^{\mathfrak{n}_{\infty}} f(i) \ \mbox{is finite}$ \\
\verb+)+

Recall that the upper limit, $\mathfrak{n}_{\infty}$, on this $\sum_{i=0}^{\mathfrak{n}_{\infty}} f(i)$, is an 
infinite positive integer constant.
\end{itemize}

\section{The Series Converges}

Use the definition, \texttt{Series-Converges-Nonstandard}, to verify, in ACL2(r), the convergence of
\[ \sum_{\mbox{\scriptsize{perfect}}(k)} \frac{1}{k} \; \; =               
  \sum_{\substack{k=1 \\ \mbox{\scriptsize{perfect}}(k)}}^{\infty} \frac{1}{k} \]
by showing this sum is finite: 
\[ \sum_{\substack{k=1 \\ \mbox{\scriptsize{perfect}}(k)}}^{\mathfrak{n}_{\infty}} \frac{1}{k} \]
Recall $\mathfrak{n}_{\infty}$ is an infinite positive integer constant. \\
Verify the previous sum is finite by showing both of the summands on the right side below are finite.
\[\sum_{\substack{k=1 \\ \mbox{\scriptsize{perfect}}(k)}}^{\mathfrak{n}_{\infty}} \frac{1}{k} \; \; \; \; =
 \sum_{\substack{k=1 \\ \mbox{\scriptsize{perfect}}(k) \\ \mbox{\scriptsize{even}}(k)}}^{\mathfrak{n}_{\infty}} \frac{1}{k} \; \; \; \; +
 \sum_{\substack{k=1 \\ \mbox{\scriptsize{perfect}}(k) \\ \mbox{\scriptsize{odd}}(k)}}^{\mathfrak{n}_{\infty}} \frac{1}{k} \]
By Theorem 2, even perfect numbers, $k$, have the form $k = 2^{i}(2^{i+1}-1)$. Since $2^{i}(2^{i+1}-1) \ge 2^{i}$,
$\frac{1}{2^{i}(2^{i+1}-1)} \le \frac{1}{2^{i}}$. Induction on $n$ verifies 
$\sum_{i=0}^{n} \frac{1}{2^{i}} = 2 - \frac{1}{2^{n}}$. Thus for any positive integer, $n$, including
$n = \mathfrak{n}_{\infty}$:
\[ 0 \; \; \le
 \sum_{\substack{k=1 \\ \mbox{\scriptsize{perfect}}(k) \\ \mbox{\scriptsize{even}}(k)}}^{n} \frac{1}{k} \; \; \; \; =
 \sum_{\substack{k=1 \\ \mbox{\scriptsize{perfect}}(k) \\ k=2^{i}(2^{i+1}-1)}}^{n} \frac{1}{2^{i}(2^{i+1}-1)} \; \; \; \; \le
 \sum_{\substack{k=1 \\ \mbox{\scriptsize{perfect}}(k) \\ k=2^{i}(2^{i+1}-1)}}^{n} \frac{1}{2^{i}} \; \le \;
 \sum_{i=0}^{n} \frac{1}{2^{i}} \; = \; 2 - \frac{1}{2^{n}} \; < \; 2 \]
By Theorem 3, odd perfect numbers, $k$, have the form $k = p^{i} m^{2}$. Since $p^{i} m^{2} \ge m^{2}$,
$\frac{1}{p^{i} m^{2}} \le \frac{1}{m^{2}}$. 
By Theorem 4, no square, $m^{2}$, appears more than once in  
\[\sum_{\substack{k=1 \\ \mbox{\scriptsize{perfect}}(k) \\ k=p^{i}m^{2}}}^{n} \frac{1}{m^{2}}\]
Induction on $n$ verifies $\sum_{m=1}^{n} \frac{1}{m^{2}} \le 2 - \frac{1}{n}$,
Thus for any positive integer, $n$, including $n = \mathfrak{n}_{\infty}$:
\[0 \; \; \le 
 \sum_{\substack{k=1 \\ \mbox{\scriptsize{perfect}}(k) \\ \mbox{\scriptsize{odd}}(k)}}^{n} \frac{1}{k} \; \; \; \; =
 \sum_{\substack{k=1 \\ \mbox{\scriptsize{perfect}}(k) \\ k=p^{i}m^{2}}}^{n} \frac{1}{p^{i}m^{2}} \; \; \; \; \le
 \sum_{\substack{k=1 \\ \mbox{\scriptsize{perfect}}(k) \\ k=p^{i}m^{2}}}^{n} \frac{1}{m^{2}} \; \le \;
 \sum_{m=1}^{n} \frac{1}{m^{2}} \; \le \; 2 - \frac{1}{n} \; < \; 2 \]
Therefore, for any positive integer, $n$, including $n = \mathfrak{n}_{\infty}$:
\[0 \; \; \le
 \sum_{\substack{k=1 \\ \mbox{\scriptsize{perfect}}(k)}}^{n} \frac{1}{k} \; \; \; \; =
 \sum_{\substack{k=1 \\ \mbox{\scriptsize{perfect}}(k) \\ \mbox{\scriptsize{even}}(k)}}^{n} \frac{1}{k} \; \; \; \; +
 \sum_{\substack{k=1 \\ \mbox{\scriptsize{perfect}}(k) \\ \mbox{\scriptsize{odd}}(k)}}^{n} \frac{1}{k} \; \; < \; \; 2 + 2 = 4 \]
and
\[\sum_{\substack{k=1 \\ \mbox{\scriptsize{perfect}}(k)}}^{\mathfrak{n}_{\infty}} \frac{1}{k} \; \; \; \mbox{is finite}. \]

The heart of this proof is that the partial sums 
\[ \sum_{\substack{k=1 \\ \mbox{\scriptsize{perfect}}(k)}}^{n} \frac{1}{k} \]
are bounded above (by 4). This can be stated and carried out entirely in ACL2.
The \textbf{Reals} and ACL2(r) are required to formally state and prove the series converges.

\appendix
\section{ACL2(r) Books}
 \subsection{\texttt{prime-fac.lisp}}
    Unique Prime Factorization Theorem for Positive Integers. \\
    An ACL2 book as well as an ACL2(r) book.
 \subsection{\texttt{perfect.lisp}}
    Perfect Positive Integers. \\
    An ACL2 book as well as an ACL2(r) book. \\
    Over 500 events, incrementally built Summer 2013 -- Spring 2015.
 \subsection{\texttt{series1.lisp}}
    The CLASSICAL series, Ser1, converges (to a STANDARD real L).
 \subsection{\texttt{series1a.lisp}}
    The CLASSICAL NONNEGATIVE series, Ser1a, converges (to a STANDARD real L).
 \subsection{\texttt{sumlist-1.lisp}}
    Some nice events from sumlist.lisp plus additional events.
 \subsection{\texttt{sum-recip-e-perfect.lisp}}
    The sum of the RECIPROCALS of the EVEN PERFECT positive integers converges.
 \subsection{\texttt{sum-recip-o-perfect.lisp}}
    The sum of the RECIPROCALS of the ODD PERFECT positive integers converges.
 \subsection{\texttt{sum-recip-perfect.lisp}}
    The sum of the RECIPROCALS of the PERFECT positive integers converges.

\nocite{*}
\bibliographystyle{eptcs}
\bibliography{perfect-paper1}
\end{document}